# Unprecedented reach and rich online journeys drive hate and extremism globally

Richard Sear and Neil F. Johnson

*Dynamic Online Networks Laboratory, George Washington University, Washington, D.C., 20052*

**Hate and extremism**[1–5] **cannot be controlled globally without understanding how they operate at scale. Both have escalated dramatically during the Israel-Hamas and Ukraine-Russia wars.**[6,7] **Here we show how the online hate-extremism system is now operating at unprecedented scale across 26 social media platforms of all sizes, audience demographics, and geographic locations; and we analyze individuals' journeys through it. This new picture contradicts notions of rabbit-hole activity at the fringe of the Internet. Instead, it shows that hate-extremism support now enjoys a direct link to more than a billion of the general global population, and that newcomers now enjoy a rich variety of online journey experiences during which they get to mingle with experienced violent actors**[8]**, discuss topics from diverse news sources, and learn to collectively adapt in order to bypass platform shutdowns. Our results mean that law enforcement**[9] **must expect future mass shooters to have increasingly hard-to-understand online journeys; that new E.U. laws**[10,11] **will fall short because the combined impact of many smaller, lesser-known platforms**[12–14] **outstrips larger ones like Twitter; and that the current global hate-extremism infrastructure will become increasingly robust in 2024 and beyond. Fortunately, it also reveals a new opportunity for system-wide control akin to adaptive vs. extinction treatments for cancer.**[15]

It is well known from the physical, engineering, biological, and medical sciences that having an accurate understanding of a complex system's infrastructure and operational dynamics is a prerequisite for any future control [16–21]. Despite the wealth of valuable literature on hate-extremism online [1–5,9,12–14,22–52], we lack any system-wide understanding of its online machinery at scale, and hence its strengths and weaknesses. Adding to this concern, elections in 2024 [53,54] will involve 2 billion voters across more than 60 countries including the U.S., India, and United Kingdom -- and there is a simultaneous explosion of new video-based social media platforms beyond YouTube (e.g. Rumble, BitChute) that can host content from video-producing generative AI [55–57]. All this will add further fuel to the religion, race, and gender hate narratives that already harm many people psychologically every day [58] and that have already incited real-world attacks including mass shootings.

Here we present the first picture of what drives hate and extremism globally across 26 social media platforms of all sizes, audience demographics, and geographic locations – and we examine individuals' journeys through it. We show that global hate-extremism now enjoys instantaneous, direct access to (and likely direct influence on) more than a billion people in the global population and that it will become increasingly robust going into 2024's global elections and beyond; that it offers a rich variety of recruitment experiences online which means that future offline attackers (e.g. shooters) will emerge from increasingly hard-to-understand pathways; that new legislation (e.g. E.U.) will not address the problem; and that current notions need changing about a niche activity at the 'fringe' of the Internet in which individuals follow some lonely path down a rabbit hole. It also reveals a new opportunity for system-wide control akin to adaptive cancer treatments.

We use a semi-automated snowballing approach [59,60] to map out the built-in social media communities that feature hate and extremism (e.g. VKontakte Club; a Facebook Page; a Telegram Channel; a Gab Group) and the links that they create between themselves, timestamped at the intra-day scale (details in SI Sec. 1). We focus on built-in communities because prior research confirms that people join these to develop their shared interests [61–64] -- and hate and extremism are no different. We go beyond Refs. [59,60] by including 26 platforms, including new video, decentralized [65], blockchain, and gaming-related platforms [66] such as Discord. Each community contains anywhere from a few to a few million users and is unrelated to network community detection. Our subject matter experts label each community as "hate" if 2 or more of its 20 most recent posts include U.S. Department of Justice-defined hate speech or promote Fascist ideologies or regime types (i.e. extreme nationalism and/or racial identitarianism) [59]. Together, these hate communities form the "hate-extremism core" (Fig. 1, center column). "Vulnerable mainstream" communities are ones that are not in this core but that were linked to directly by a hate community (Fig. 1, right column). Vulnerable mainstream communities' views can vary significantly, but they mostly represent a benign population that have become targets of the hate-extremism core (SI Sec. 1.2). Nuancing these definitions would not change our system-level picture significantly. A link to community B can appear in community A at some time $t$ if A creates a post that contains a URL linking directly or to content posted in B (e.g. SI Figs. S1, S4, and S8 for Hamas). The link directs the attention of A's members to B, which might be on a different platform and/or another language. A's members can then add content on B without B's members knowing about this inbound link. Hence B's members can unwittingly experience direct influence and exposure from A's narratives. In this study, we do not include links originating in vulnerable mainstream nodes. Only public communities are accessed, but this ecosystem provides a skeleton on which private communities grow and communicate (SI Fig. S3). Individual avatar pseudonyms were extracted from VKontakte for Fig. 2. At the time, these were openly available, but no private information was extracted. We know what happened to a particular pseudonym when referred to by other members of their community in open discussions [8]. We quantify hate types (Fig. 3) using a machine learning classifier with a high degree of accuracy as measured against human subject matter experts (see Ref. [59] for full details).

Figure 1 shows the new picture that emerges of the global hate-extremism ecosystem. It features highly interconnected communities within and across platforms together with a *direct* connection to more than a billion members of the global population (i.e. 1 click away from the hate-extremism core). There are 1,592 hate communities (nodes) totaling roughly 50M individuals. This number is estimated by multiplying the average number of members per community by the number of communities on each platform shown in Fig. 1, center. Over 3.5 years, these hate communities have created 365,424 links between themselves and 4,015,141 links into 490,643 vulnerable mainstream communities.

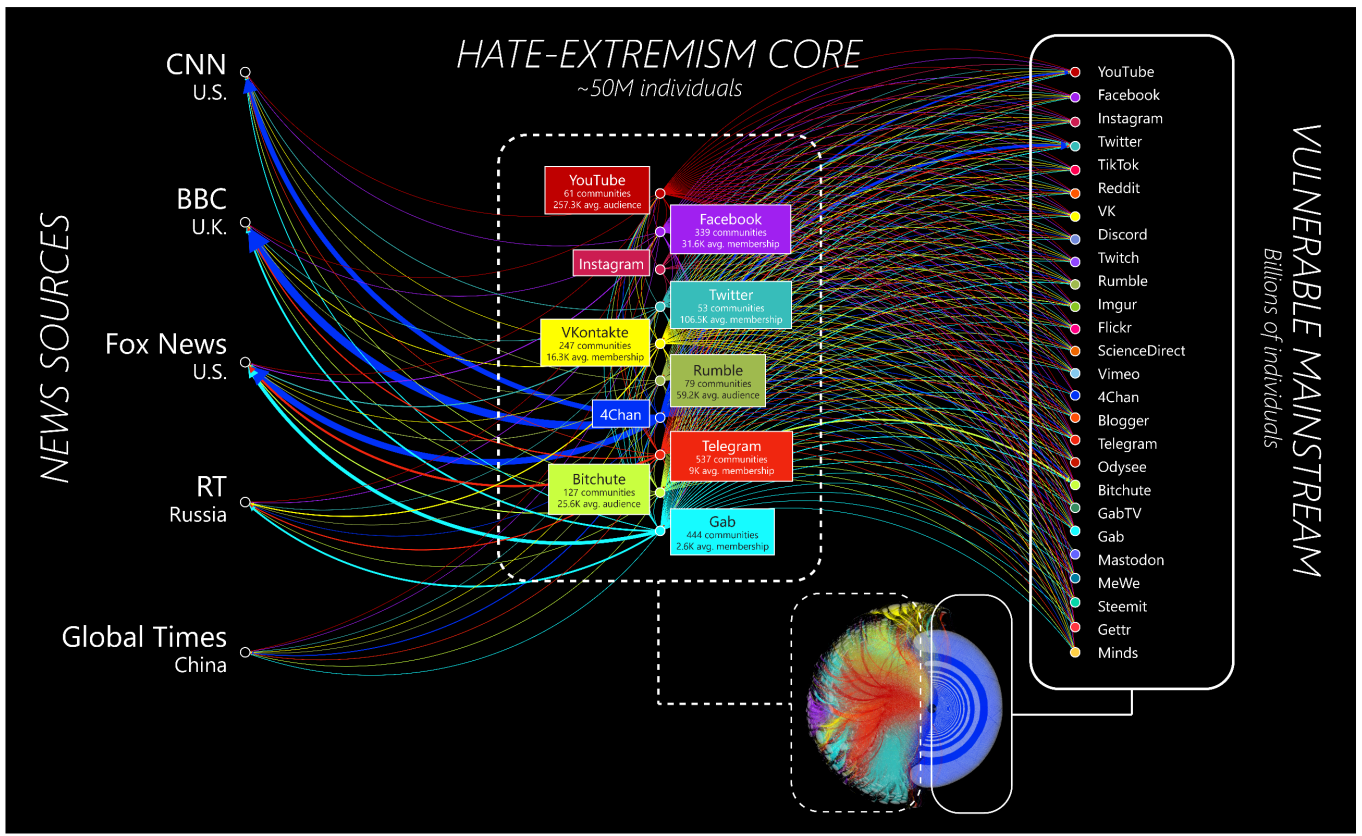

**Figure 1: Hate-extremism ecosystem across platforms, languages, and countries.** Full network data is available online but for visual convenience the center and right columns show links between hate communities aggregated over time, and hate communities aggregated within each platform. Weighted links from hate communities (hate-extremism core, dashed box) to news sources (left column) indicate number of times hate communities link to news. Weighted links to mainstream targets (i.e. to vulnerable mainstream communities, solid box on right) indicate number of links and hence imply level of direct influence of hate-extremism core. Lower circular diagram is disaggregated version (ForceAtlas2 layout). Each colored dot (node) is a single hate community; links have color of the source node. Each white dot (node) is a vulnerable mainstream community. Some vulnerable mainstream communities have far higher exposure than others: they attract more links into them, and so sit closer to hate-extremism core because of node-repulsion-link-attraction ForceAtlas2 layout. Red (Telegram) regions feature frequent links into Hamas' primary Telegram community (SI Fig. S8). See SI Section 2.1 for more details about this figure.

This new picture challenges current perceptions in multiple ways. First, the core of approximately 50 million individuals in hate communities now has direct online access to more than a billion vulnerable mainstream community members and hence more than 1/8 of the global population [67]. It is not a fringe activity sitting at some 'distance' from the mainstream. Second, hate communities on each of the platforms listen to a diverse spectrum of news sources including CNN and the BBC, i.e. it is not a far-right news echo chamber as often claimed. Such diversity may help tempt new recruits wary of echo chambers. Third, the biggest platforms do not dominate the ecosystem. Instead, all platforms down the global size ranking are roughly equivalent in terms of connectivity – which means the new E.U.

regulations' large platform focus will likely be ineffective. The new picture also reveals routes being constructed to bypass larger platforms' mitigations (e.g. Fig. 2A). Fourth, video channels which can feature AI-altered videos (e.g. Rumble, Bitchute, YouTube) play a key role. Fifth, Twitter does not stand out as a major player, despite being the focus of most academic studies. Sixth, the resulting dynamical network is unlike other real-world networks that we know of, and is unlike Erdos-Renyi (random), small-world, and scale-free network idealizations. For example, instead of a single giant component, there are typically several large components which change size and membership over time, with each comprising communities (nodes) from many different platforms. Seventh, the number and identities of the hate communities is surprisingly steady over the scale of years which suggests they will also dictate its behavior in 2024 and well beyond. The number of links increases massively every day (and of course some past links will be forgotten or lost). This happens at a reasonably steady rate, so the trends are likely to continue. This means the ecosystem's networks are continually strengthening, and hence so will the strength of their direct influence on the global mainstream (Fig. 1).

Eighth and arguably most importantly in terms of future real-world attacks, the hate-extremism ecosystem (Fig. 1) now offers newcomers a rich variety of online journey experiences during which they get to mingle with experienced violent actors and learn to collectively adapt in order to bypass moderator shutdowns (e.g. SI Fig. S9). This calls into question the notion that any particular individual follows some lonely path down a rabbit hole toward radicalization and perhaps a real-world attack. Figure 2 shows that even for just the limited topic of anti-U.S. hate on a single platform (VKontakte), there is a wide variety in the length of individuals' journeys through the hate-extremism ecosystem. In less than 6 months, each individual joined between 1 and 403 hate communities comprising a mix of violent hate (green) and extremely violent hate (red) communities. There is no 'typical' pathway, which is consistent with the lack of such pathways being found by law enforcement following most (if not all) recent shootings globally. Moreover, it means law enforcement should expect increasingly hard-to-understand paths by mass shooters as the ecosystem evolves in the future. The examples shown (individuals B-E) are some of the many that later became violent real-world attackers.

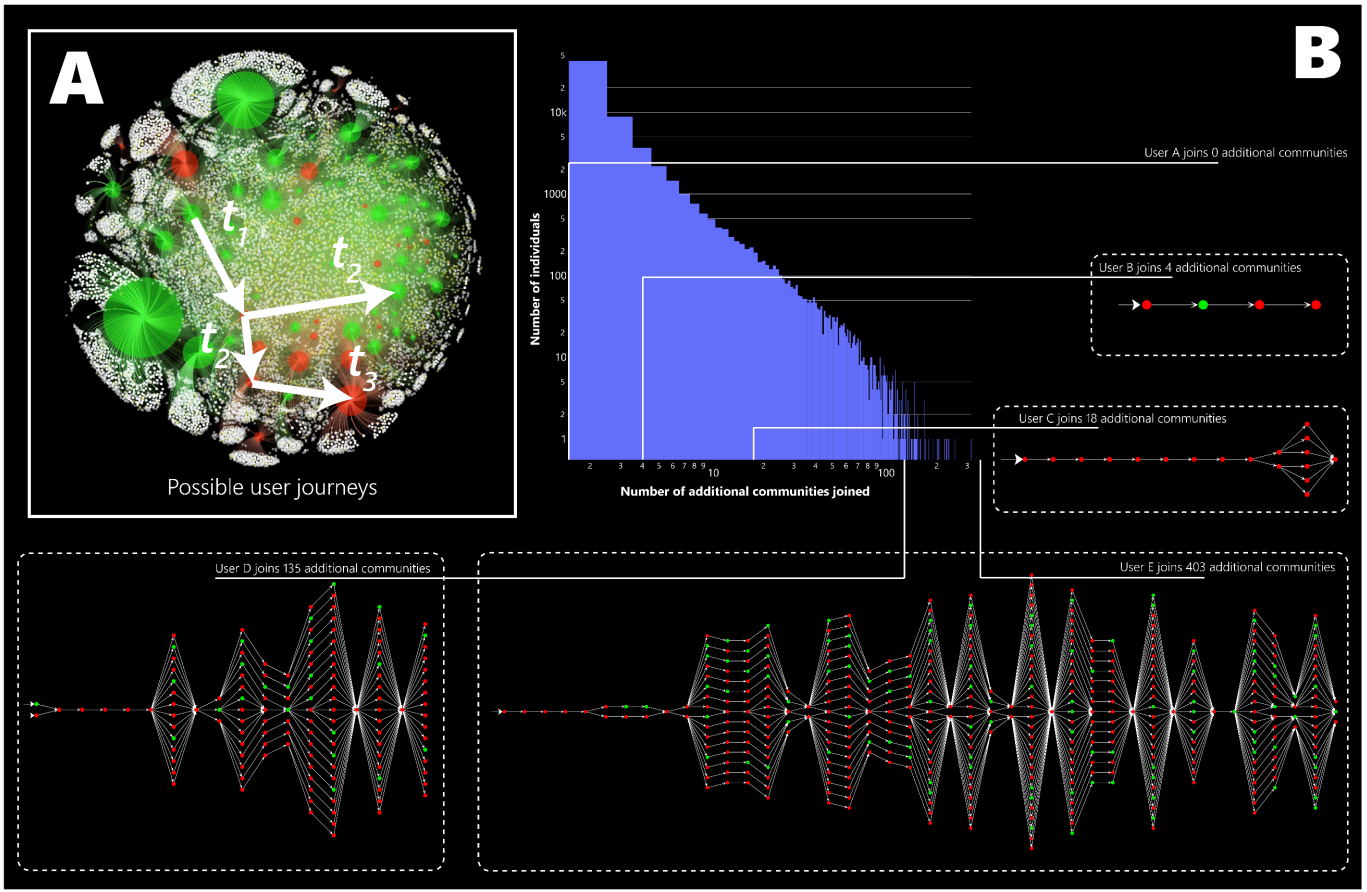

**Figure 2: Online journeys that individuals take through hate communities in the hate-extremism ecosystem.
A.** Illustration shows anti-U.S. hate communities on VKontakte. Red circles are anti-U.S. hate communities that were so violent, they were later banned. Green circles are anti-U.S. hate communities that were violent but less so, and so were not banned at the end of the study period. Small white circles are individuals. **B.** Histogram of actual journeys through subsystem in A. Examples of journey timelines (i.e. community joining) of some individuals. Each individual was later killed in their real-world role as a violent attacker (e.g. while in Syria fighting U.S. troops). See SI Sec. 2.2 for additional details about this figure.

This new picture of a highly dynamic, globally interconnected infrastructure that listens to many types of news sources suggests the ecosystem should be able to respond at the global scale almost instantly to world events. Figure 3 provides evidence of this. There is a huge and almost instantaneous rise in antisemitic hate in the minutes following Hamas' attack on October 7, 2023 (first vertical line) -- and to a much lesser extent Islamophobia -- despite the fact that Hamas had just attacked Israel and Israel had yet to respond. Also, new recruits to key Hamas communities on Telegram instantly increased while new recruits to Hezbollah rose more slowly over the subsequent days (SI Fig. S10). By contrast, there were only minor increases following the Ahli Arab Hospital explosion (second vertical line) which suggests that the hate-extremism ecosystem had already factored in its amplified antisemitism.

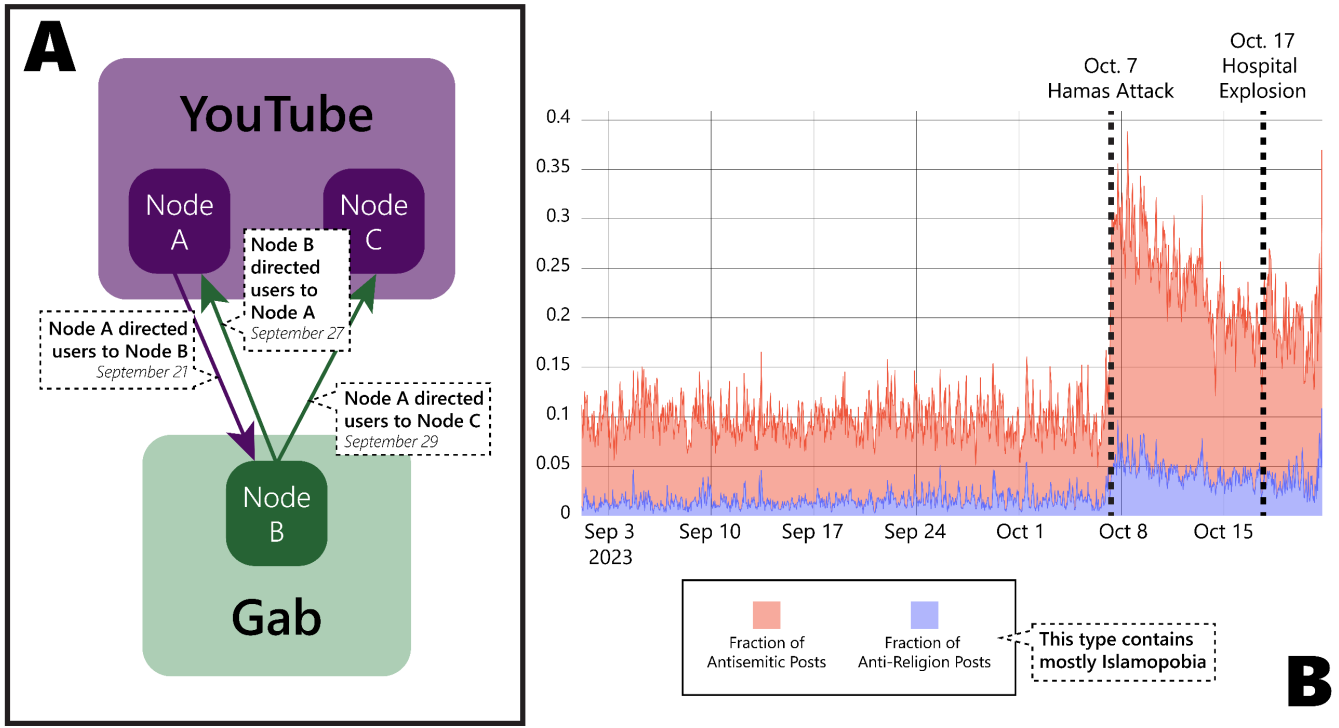

**Figure 3:** A. One example of the ubiquitous inter-platform bypasses created by inter-platform links, that was in place just prior to the Hamas attack. These enable hate and extremism to temporarily move from a particular platform to a different platform before returning, in order to avoid sudden moderator attention on the first platform. This makes it appear to the first platform's moderators as though they had eradicated this particular hate-extremism (e.g. Node A), only to later see it re-appear out of the blue. This offers a resolution of the common dichotomy whereby a platform says it is actively undertaking mitigation yet users on it see no change. B. Right panel: antisemitic hate level shows huge instantaneous increase in the minutes following Hamas' attack on October 7, 2023 (first vertical line), and to a far lesser extent so does Islamophobia. Only minor increases follow the Ahli Arab Hospital explosion (second vertical line). See SI Sec. 2.3 for additional details about this figure.

The strong entanglement between all platforms (Fig. 1) with its numerous inter-platform bypasses (Fig. 3A) means that moderation and legislation (e.g. E.U.) focused on the so-called major platforms (e.g. Twitter, Facebook, YouTube) are unlikely to work. Instead, and akin to new adaptive treatments for cancer [15], this strong inter-platform mixing suggests a new system-wide control be adopted in which policymakers aim for system-wide containment guided by the observed activity across Fig. 1, as opposed to seeking costly and elusive total extinction. We demonstrate this by adapting Epstein and MacKay's equations [68,69] to describe the mutual attrition between hate-extremism activity $H(t)$ and the activity of moderators with finite resources $M(t)$ where $H(0) > M(0)$ (hate-extremism currently winning) in this limit of strong inter-platform mixing. If $\frac{\partial H}{\partial t} = -mM$ and $\frac{\partial M}{\partial t} = -hH$ where the fighting efficiencies $m$ and $h$ are constants, the exact solution shows (see SI Sec. 6) that moderators will only be able to drive hate-extremism $H(t)$ to extinction if $\sqrt{\frac{m}{h}} > \frac{H(0)}{M(0)}$. For example, $H(0) = 4M(0)$ means that moderators' fighting-efficiency $m$ will have to be more than sixteen times the hate-extremism fighting-efficiency $h$. If $\frac{\partial H}{\partial t} = (-mM)H$ and $\frac{\partial M}{\partial t} = (-hH)M$, then $m$ only needs to be more than four times $h$. In both these symmetric scenarios, $m$ must be significantly larger than $h$ in order for moderators to drive hate-extremism to extinction. If instead moderators use an adaptive approach to 'ambush' multi-platform activity as it arises in Fig. 1, the battle become asymmetric: $\frac{\partial H}{\partial t} = (-mM)H$ and $\frac{\partial M}{\partial t} = -hH$. Then, even if fighting-efficiencies are the same ($m = h$), $M(0)$ moderators can achieve stalemate and hence containment of a hate-extremism ecosystem of size $M^2(0)/2$. This means 10,000 moderators can contain the actual hate-extremism core shown in Fig. 1 of 50 million individuals. This crude calculation can be made more realistic by adding platform heterogeneity [70], but the same general conclusions will arise in the well-mixed limit.

**Supplementary Information (SI)**
Contents:

1. Methodology
    1.1 Data collection
    1.2 Discussion of vulnerable mainstream communities

2. Additional details on figures in main paper
    2.1 Hate-extremism network
        2.1.1 Aggregated visualization
        2.1.2 Force-directed network
    2.2 Individual pathways
        2.2.1 Force-directed network
        2.2.2 Pathway diagrams
    2.3 Adaptive network structure primes users to react quickly to current events

3. Location in hate-extremism core of content related to global conflicts

4. Coordinated rearrangement of online networks

5. Membership of Hamas and Hezbollah groups before and after October 7, 2023

6. Proof of quoted mathematical results

**Funding:** N.F.J. is supported by U.S. Air Force Office of Scientific Research awards FA9550-20-1-0382 and FA9550-20-1-0383, and The John Templeton Foundation.

**Authors contributions:** R.S. collected edge data, managed the databases, and developed software. N.F.J. supervised the project, performed the modeling, and wrote the paper drafts.

**Competing interests:** The authors have no competing interests, either financial and/or non-financial, in relation to the work described in this paper.

**All correspondence and material requests should be addressed to** N.F.J. neiljohnson@gwu.edu

**Data and materials availability:**

**Data Availability**
The datasets used in this study contain sensitive information from social media platforms. To comply with data protection standards and avoid potential misuse, all the raw text content cannot be shared publicly; however, the preprocessed derivative datasets that can be used to reproduce the results in the study are available in our Data Access repository, https://github.com/gwdonlab/data-access. This provides readers with access to the minimum dataset that is necessary to interpret, verify, and extend the research in this article.

**Code Availability**
The code used was standard Python libraries for web crawling (e.g. BeautifulSoup, lxml), quantitative analysis (e.g. pandas, numpy), and data visualization (Plotly), all of which are open and free. Gephi (also free and open-source) was used to produce network visualizations. Mathematica was also used for quantitative analysis and Adobe Illustrator was used to produce the final figures. These are well-known commercial products available through site licenses in many universities.